\documentclass[useAMS,usenatbib,usegraphicx]{mn2e}

\title[Debris discs around nearby Solar analogues]{Debris discs 
around nearby Solar analogues}
\author[J. S. Greaves et al.]{J. S. 
Greaves$^{1}$\thanks{E-mail: jsg5 at st-andrews.ac.uk}, M. C. 
Wyatt$^{2}$ \& G. Bryden$^{3}$\\
$^{1}$SUPA, Physics \& Astronomy, University of St Andrews, North 
Haugh, St Andrews, Fife KY16 9SS, UK\\
$^{2}$ Institute of Astronomy, University of Cambridge, 
Cambridge CB3 0HA, UK\\
$^{3}$ Jet Propulsion Laboratory, California Institute of 
Technology, 4800 Oak Grove Drive, Pasadena, CA 91109, USA
}
\begin{document}

\date{Accepted 2009. Received 2009; in original form 2009} 
\pagerange{\pageref{firstpage}--\pageref{lastpage}} \pubyear{2006}

\maketitle

\label{firstpage}

\begin{abstract}

An unbiased search for debris discs around nearby Sun-like stars
is reported. Thirteen G-dwarfs at 12-15 parsecs distance were
searched at 850~$\umu$m wavelength, and a disc is confirmed around
HD~30495. The estimated dust mass is 0.008~M$_{\oplus}$ with a net
limit $\la 0.0025$~M$_{\oplus}$ for the average disc of the other
stars. The results suggest there is not a large missed population
of substantial cold discs around Sun-like stars -- HD~30495 is a
bright rather than unusually cool disc, and may belong to a few
hundred Myr-old population of greater dust luminosity. The
far-infared and millimetre survey data for Sun-like stars are well
fitted by either steady state or stirred models, provided that
typical comet belts are comparable in size to that in the Solar
System. 

\end{abstract}


\section{Introduction}

Debris discs represent the fallout of collisions of comets or
asteroids around main-sequence stars. Showers of dust particles
are continually regenerated, and these orbiting grains produce
thermal emission from the mid-infrared to the millimetre,
depending on the distances of the particles from the host star
and so their equilibrium temperatures.  The Sun's Kuiper Belt of
comets would be a very faint example of the phenomenon if viewed
externally, but much brighter exo-systems are known, which must
be supported by belts of colliding bodies that are more populated
and/or more perturbed. 

\begin{table*}
 \centering
 \begin{minipage}{160mm}
  \caption{Star and disc properties. The mean and 1-$\sigma$ 
error from 3$\sigma$-clipped photometry at 850~$\umu$m are quoted 
except for the summed flux of HD~30495. Dust masses and 
$3\sigma$ limits are for 60~K grains with emissivity of 1 
cm$^2$/g; HD~72905 is scaled from HD~30495 assuming the 
millimetre flux-ratio equals the 70~$\umu$m excess-flux ratio of 0.25 
\citep{trilling}. HD~172051 is background-contaminated (see text) 
so no mass is given. Spectral type and multiplicity information 
are from the NStars database, and metallicities  (log-abundance 
of iron versus Solar) are from \citet{vf}. Stellar ages are 
from \citet{barnes} or via the same gyrochronology method using 
B-V and estimated rotation periods from \citet{wright}; bracketed 
values of 0.5~Gyr are for candidate members of the UMa moving 
group \citep{king}. 
}
  \begin{tabular}{rccccccl@{}}
  \hline
star 	 & flux	& dust mass			& type	& distance & age 	& [Fe/H] & notes \\
(HD)     & (mJy)& (M$_{\oplus}$)		&	& (pc)     & (Gyr)	&        &	 \\
  \hline
30495	 & 5.6 $\pm$ 1.7	& 0.008		& G3 V	& 13.3	& 0.45		& +0.01	& ISO \& Spitzer disc \\
34411	 & +0.6 $\pm$ 2.2  	& $\leq$ 0.009	& G0 V	& 12.6	& 4	 	& +0.12	& \\
72905	 & +1.7 $\pm$ 1.7	& $\sim 0.002$*	& G1.5 V& 14.3	& 0.15(--0.5)	& ---	& Spitzer disc (*mass if dust at 60~K) \\
140538	 & --1.1 $\pm$ 1.4	& $\leq$ 0.007	& G5 V	& 14.7	& 4		& +0.08	& m$_V$ = 12 companion at $4''$ \\
141004	 & +1.7 $\pm$ 2.4	& $\leq$ 0.008	& G0 V	& 11.8	& 3--6		& +0.05	& \\
144579	 & --1.8 $\pm$ 1.2	& $\leq$ 0.006	& G8 V	& 14.4	& 6		& --0.69& m$_V$ = 14 companion at 70$''$ \\
146233	 & --1.5 $\pm$ 2.1	& $\leq$ 0.010	& G1 V	& 14.0	& 4		& +0.03	& `Solar twin'\\
                                                                          	&&&&&&& \citep{st} \\
147513	 & +0.8 $\pm$ 2.0	& $\leq$ 0.008	& G3/5 V& 12.9	& (0.5--)0.6	& +0.09	& planet of $\geq$ 1 M$_{Jup}$ \\
        	          			                  	  	&&&&&&& \citep{mayor} \\
                	 			       	                  	&&&&&&& white dwarf companion at 345$''$ \\
157214	 & --1.2 $\pm$ 1.2	& $\leq$ 0.006	& G0 V	& 14.4	& 4		& --0.36& \\
160269	 & +0.8 $\pm$ 1.7	& $\leq$ 0.008	& G0 V	& 14.1	& (0.5?)	& ---	& triple: K3 V, M1 V at 1.5, 740$''$ \\
172051	 & (+7.6 $\pm$ 2.1)	& ---	        & G5 V	& 13.0	& 4		& --0.26& \\
176051	 & --0.7 $\pm$ 1.4	& $\leq$ 0.008	& G0 V	& 15.0	& 2.5		& ---	& m$_V$ = 8 companion at 1.2$''$ \\
196761	 & +1.5 $\pm$ 1.3	& $\leq$ 0.007	& G8/K0 V& 14.6	& 5		& -0.28	& \\
\hline
75732	 & $\sim$ 0 $\pm$ 0.4	& $\leq$ 0.002	& G8 V	& 12.5	& 8		& +0.31	& SCUBA data: \citet{rayj} \\
                             				       	          	&&&&&&& 5 planets of $\geq$ 0.03--3.8 M$_{Jup}$ \\
                         	        		      	          	&&&&&&& \citep{fischer} \\ 
                                	  		  		  	&&&&&&& m$_V$ = 13 companion at 85$''$ \\
  \hline
\end{tabular}
\end{minipage}
\end{table*}

A number of (sub)millimetre surveys have looked at nearby Solar
analogues to see if massive belts of comets could be common
\citep{lindroos,holmes,liu,carpenter,najita,williams}. The
debris is expected to be cool if these belts lie at tens of AU
outwards and are heated by stars of $\sim 1$~L$_{\odot}$, so
long wavelength emission is expected. Discs were detected with
moderate frequency for stars a few hundred Myr old, but no new
detections were made for ages comparable to the Sun's 4.5 Gyr.
However, the sampling of mature stars is poor and the larger
studies have only reached sensitivities of a few mJy around 1~mm
\citep{sunss}. 

Thus, the prevalence of cool dust around Solar analogues remains
unknown. Of particular interest are a few examples of discs of
Sun-like stars with dust thought to be cooler than 40~K
\citep{holmes,najita,lindroos,liseau}. These could have escaped
detection in the recent Spitzer 24--70~$\umu$m surveys that have
discovered new debris discs around Sun-like stars, as such cold
dust has low contrast with the stellar photosphere. Large cool
discs are interesting as they may still be evolving --
\citet[Eq.~41]{kb08} predict that for discs exceeding 200~AU in
radius, 1000~km bodies would form later than the Sun's present
age of 4.5~Gyr, and these could create late debris showers by
perturbing the comet belt. Debris is in fact seen in systems as
old as 10~Gyr, in the nearby case of $\tau$~Ceti
\citep{tauceti}. 

We have thus searched a small sample of local Solar analogues to
look for cold debris. This is the first such unbiased survey in
the submillimetre -- based only on distance, without
pre-selection from stellar age, multiplicity, or evidence of
far-infrared excess -- and is a pre-cursor to the SCUBA-2 Legacy
Survey of nearby stars \citep{sunss}. Here we present the results 
on masses of cold dust around nearby G-dwarfs and make some 
comparisons with recent models.

\section{Survey data}

The SCUBA camera on the 15 m James Clerk Maxwell Telescope was
used to survey G-dwarf (luminosity class V or IV-V) stars at
distances of 10-15~pc. The initial selection was made from the
NStars Spectra online catalogue\footnote{http://nstars.nau.edu/,
http://stellar.phys.appstate.edu/}, yielding 29 stars.
Eliminating 4 objects unsuitable for the JCMT in Hawaii (with
Dec. $< -40^{\circ}$) and 11 sources clashing in scheduling with
a cosmology program (RA around 0--4, 9--15h) left 14 targets.
All of these were observed except for HD~75732 (55~Cnc) which
already has a deep SCUBA limit \citep{rayj}. The distance
range\footnote{At closer distances, there are 12 further
G-dwarfs, but these were omitted because any large discs would
overfill the effective photometric beam. These stars include one
known debris disc, HD~10700 ($\tau$ Ceti), and debris candidate
HD~131156 \citep{holmes,decin}.} of the final 13 targets was
actually narrower, only 12--15 pc, making very uniform
sensitivity to dust mass possible.

Because the survey is unbiased apart from co-ordinate bounds,
the stars have typical G-dwarf properties (Table~1). Six of the
stars are in multiple systems where any comet belts
could be perturbed \citep{trillbin}, and two have known giant
planets. The stellar metallicities are one-fifth to twice Solar
and the stellar ages span the main-sequence range for G-dwarfs.
These ages adopt the accurately calibrated gyrochronology
(spin-down) method of \citet{barnes} with most having similar
estimates from isochrones or chromospheric activity
\citep{takeda,wright,nordstrom}. Most of the stars are a few Gyr
old, with a slight enhancement of sub-Gyr objects compared to a
random population with $\sim 10$~Gyr lifetimes, due to the
existence of local associations. In particular, HD~160269, 72905
and 147513 are possible members of the 0.5~Gyr-old Ursa Major
moving group \citep{king}. 

\subsection{Observation methods}

The stars were observed in extended photometry mode
\citep{sheret}, which is a compromise in observing speed between
staring at one point and full mapping, for a detector array
which undersamples the sky. The central SCUBA 850~$\umu$m
bolometer observed a pattern of points around each star, at
offsets of $1''$ and $7.5''$. With the $15''$ (full-width
half-maximum) diffraction-limited beam, this mode collects a
significant fraction of the flux from a disc extending up to
$15''$ from the star, i.e. up to approximately 200~AU radius for
stars at 12--15~pc distance. 

Integration times were 1.4--3.4 hours per star, summing to 36
hours and requiring nearly 70 hours at the telescope. The
observing campaign ran from 2003 July to 2005 May, over 17
nights. All of the stars except two (HD 34411 and 146233) had
observations on more than one night. Good to moderate conditions
were used, with zenith sky opacities at 850~$\umu$m of 0.23 to
0.52. Data reduction was performed using the oracdr pipeline
\citep{oracdr} plus custom IDL routines \citep{etacorvi},
including extinction corrections from skydips on each night, and
applying a flux conversion factor of 307~Jy/V per beam (with a
$1 \sigma$ spread of $\pm 34$~Jy/V per beam over the 13 nights
with FCF measurements). This conversion factor was derived from
point-like calibrator sources observed using the same extended
photometry mode, and is higher than for normal on-source
photometry, as two-thirds of the time is spent looking at the
half-power points of the calibrator. For any extended discs, 
the true flux will be underestimated, but less so than for 
normal photometry because of the greater spatial coverage
\citep[Table~3]{sheret}. 

\section{Results}

\begin{figure}
\label{figure1}
\includegraphics[width=70mm,angle=0]{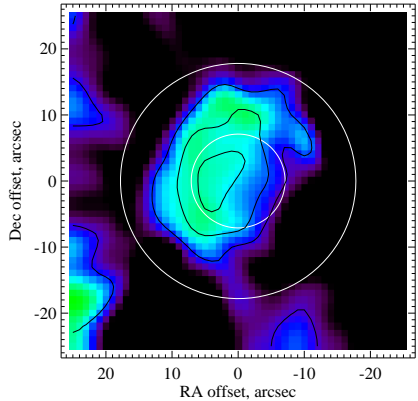}
 \caption{Disc image for HD~30495. Colour scale shows the mean 
of all measurements within $7''$ of the pixel, further smoothed by 
a $4''$ Gaussian. Black contours represent corresponding 
signal-to-noise at levels of 1, 2 and 3. The inner white circle 
delineates the beam of the central bolometer, and points from 
here to the outer circle are interpolations, as the first ring 
of bolometers samples only from this radius, i.e. $\geq 18''$. } 
\end{figure}

The fluxes are listed in Table~1, and each data set was also
examined as an image to search for any disc-like structure. No
new discoveries were made among the 13 stars observed, but the
ISO and Spitzer excess of HD~30495 is confirmed as a debris disc
by SCUBA (Figure~1). The total photometric flux is $5.6 \pm
1.7$~mJy (assuming a point-like morphology), to which the
photosphere contributes a negligible $\approx 0.15$~mJy. The disc
is oriented at a position angle of approximately $-12^{\circ}$,
and adding up all data points in two quadrants around this major
axis and within $10''$ of the star gives a flux of $4.9 \pm
1.2$~mJy, compared to only $0.7 \pm 1.2$~mJy for the two
minor-axis quadrants. We infer that the disc is seen rather
edge-on, and is also marginally extended as the flux density at
$7''$ along the major axis is $4.8 \pm 2.0$~mJy/beam, similar to
$4.3 \pm 2.0$~mJy/beam for the $1''$ points. The under-sampled 
data (Figure~1) may not capture all the disc flux, but the 
estimate of $5.6 \pm 1.7$~mJy is roughly consistent with the 
flux pattern over the quadrants. 

The interpolated image of HD~30495 suggests the major axis covers
up to $25''$ diameter, which after deconvolution from the $15''$
beam implies dust extends out to about $10''$ or 130~AU radius.
In resolved images, $r_{\rm disc} \sim 100-150$~AU is typical for
F, G and K stars \citep[e.g.]{kalas}. The rather edge-on
orientation is consistent with the estimated stellar inclination,
assuming the star and disc are co-planar: from the stellar
rotation period, radius and $v sin i$ \citep{gaidos,vf}, the
value of $i$ is approximately $50-75^{\circ}$. 

The spectral energy distribution for HD~30495 (Figure~2) implies
dust at approximately 57~K, similar to the 70~K estimated by
\citet{zs} based on the far-infrared SED. The disc is observed to
be larger than the radius of 20~AU inferred for a blackbody SED
\citep{zs}, suggesting some greybody emission where particles
that are inefficient emitters lie further from the star at
equivalent temperatures \citep[e.g]{sheret}. 

\begin{figure}
\label{figure2}
\includegraphics[width=70mm,angle=0]{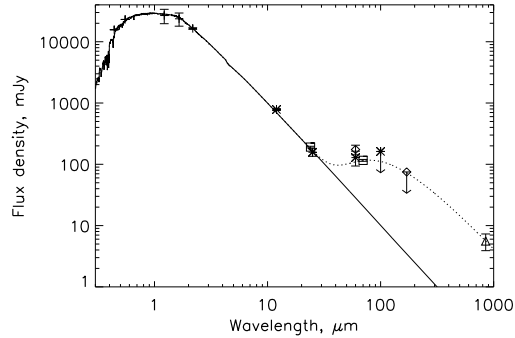}
 \caption{Spectral energy distribution for HD~30495. The 
solid line is a Kurucz model for a G3 star scaled to the 2MASS 
K-band flux, and the dotted line adds a 57~K blackbody 
with $L_{IR}/L_* = 4 \times 10^{-5}$. The + symbols are 
B, V and JHK measurements from Tycho and 2MASS; $*$ 
symbols are IRAS fluxes from SCANPI (colour corrected for 
the photospheric spectrum); diamond and square symbols 
represent ISO and Spitzer respectively. Upper limits are $3 
\sigma$.
} 
\end{figure}

\begin{figure}
\label{figure3}
\includegraphics[width=80mm,angle=0]{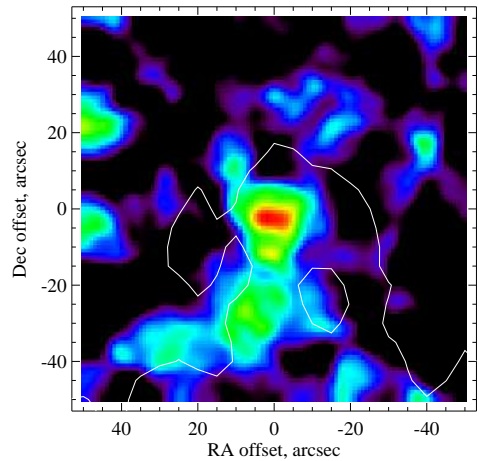}
 \caption{SCUBA image of HD172051 but in S/N units (for clarity 
in displaying unevenly-sampled data). Colour scale is from 0 to 
4, with yellow roughly at 3-sigma. The white contour (arbitrary 
level) is from a 70~$\umu$m Spitzer image of the same region. }
\end{figure}

A positive signal was also found towards HD~172051 (Figure~3).
However, the structure is not disc-like but more extended, and is
approximately matched by a contour from Spitzer 70~$\umu$m data
\citep{chas}. In this larger image, widespread emission is seen
that extends towards the Galactic Plane. Since this star is
within 15 degrees of the Galactic Centre, the signal could be 
from molecular clouds, and there is no compelling evidence
of a disc contribution to the flux at either wavelength. 

All of our target stars have now been searched with Spitzer for
debris discs, using MIPS at 24 and 70~$\umu$m wavelengths.
(Archived data for HD~140538, 141004, 160269 and 176051 are not
yet published but do not show any obvious cool excesses.) Of the
twelve objects with SCUBA limits, one has a debris detection:
HD~72905 has dust emission over wavelengths of at least 25 to
70~$\umu$m \citep{chas-irs,bryden,trilling}. This has been
interpreted as arising from warm and cool planetesimal belts,
with the SCUBA upper limit suggesting T~$\ga 40$~K for the cool
dust.

\subsection{Disc properties}

The incidence of cool moderately-massive debris discs is not
high, since we detect only one example among thirteen Sun-like
stars ($\sim 8$~\%). HD~30495 appears to be detectable in the
submillimetre mainly because the disc is bright (e.g. 95~mJy at
70~$\umu$m versus 24~mJy for HD~72905) rather than because it
has cold dust that is better matched to long wavelengths.

Table~1 lists the upper limits to debris masses, assuming an
emissivity of 1~cm$^2$/g and 60~K grains as for HD~30495
(adopting e.g. the 40~K lower limit from HD~72905 would increase
the masses by 50~\%). For the majority of the G-dwarfs, any
debris must be below a level of about 0.01~M$_{\oplus}$ or about
a lunar mass. Somewhat larger masses could exist in very distant
cold dust belts without exceeding either SCUBA or Spitzer upper
limits. 

Lower mass limits can be reached by co-adding the data for ten
fields without disc or background emission (Table~1), yielding a
net 850~$\umu$m flux of $-0.3 \pm 0.5$~mJy. Thus, no dust (or
photospheric emission) is found on average around these ten
stars. The 3$\sigma$ dust upper limit of 1.5~mJy implies that the
mean disc around these stars comprises less than
0.0025~M$_{\oplus}$, for 60~K grains and representative distance
$\approx 14$~pc. This limit is below the dust masses for nearly
all the submm-detected debris discs around Sun-like hosts
\citep{wyattrev}, while the 0.008~M$_{\oplus}$ of material around
HD~30495 is within the typical range of a few thousandths to a
tenth of an Earth-mass of dust \citep[Fig.~3]{wyattrev}. There is
a marginally positive signal of $+2.0 \pm 0.9$~mJy for all the
four sub-Gyr sources averaged together, versus a null signal of
$-0.6 \pm 0.6$~mJy for eight older objects, suggesting a possible
decline of dustiness with age, as discussed further below. 

The disc luminosities and limits here are similar to those from
far-infrared data. Converting the 850~$\umu$m fluxes to
fractional luminosities \citep[Eq.~4]{chas}, $L_{dust}/L_*$ is $4
\times 10^{-5}$ for HD~30495, similar to HD~72905 at $3 \times
10^{-5}$ \citep{chas-irs}, while the SCUBA limits for the older
stars correspond to $\la 2-6 \times 10^{-5}$.  For the co-add of
the ten blank SCUBA fields, $L_{dust}/L_*$ is $\la 1 \times
10^{-5}$ for a typical stellar effective temperature of 5800~K. 

\section{Discussion}

The SCUBA survey has shown that the debris disc population can
not be greatly boosted by previously missed cold discs of
moderate mass. Only one disc was found in an unbiased sample of
thirteen nearby Solar analogues, and the dust temperature of
around 60~K is high enough to also yield prominent far-infrared
excesses for this star \citep{habing,trilling}. Combining the new
SCUBA results with earlier millimetre surveys for debris around
FGK dwarfs \citep{lindroos,holmes,sheret,najita,williams}, in
total 33 Sun-like stars $\ga 150$~Myr old have now been observed,
spanning the epoch after terrestrial planet formation (e.g. the
latest age of Earth completion, \citep{touboul}) up to the end of
the main-sequence. There are seven debris
candiates\footnote{These stars are HD~8907, 30495, 38393, 48682,
109085, 131156 and 206893. HD~38393, 131156 and 206893 are only
at $\approx 2.5-3\sigma$ confidence in the millimetre and the
first two having ISO excesses not confirmed by Spitzer. Other
Sun-like stars from individual studies also have millimetre
debris detections: see \citet{wyattrev} for a summary.} and five
also have confirmed far-infrared excesses. The far-infrared to
millimetre fluxes have been fitted successfully with grain
temperatures ranging from 50 to 100~K, so there is no indication
of additional debris in very cold and distant comet belts for
these stars. Thus although there are some systems with evidence
of dust below 40~K \citep[e.g.]{lindroos}, this appears to be
uncommon around Solar analogues. 

Most millimetre detections have been made towards young stars.
Dividing the 33 objects into two broad age bins of width 1 dex
around 500~Myr and 5~Gyr (to reduce the effects of uncertain
dating), there are up to 6 debris detections (2 tentative) out
of 13 stars in the younger group, compared to only 1/20 in the
older set. With Poisson counting errors, the detection rate
then nominally declines from (30-45)~$\pm$~(15-20)~\% to $5 \pm
5$~\% for stars an order of magnitude older. However, the
counts are skewed because some projects targeted more young
stars and/or prior debris-candidates. Hence, the true
incidences are likely to be lower, and the trend with time may
be partly forced by deeper searches towards targets thought
more likely to yield detections. For example, four of the seven
young prior-debris candidates have possible millimetre
detections, but these were also some of the deepest
observations made in the respective surveys
\citep{holmes,sheret,williams}. In general, a decline of
dustiness with age is plausible, especially as the fraction of
prior debris candidates ($\approx 55$~\%) happens to be the
same in both the young and old samples, reducing one source of
bias. Also, the oldest millimetre detection is HD~48682 at only
$\sim 3$~Gyr, suggesting that stars several Gyr old may tend to
be millimetre-faint. 

\begin{figure}
\label{figure4}
\includegraphics[width=60mm,angle=270]{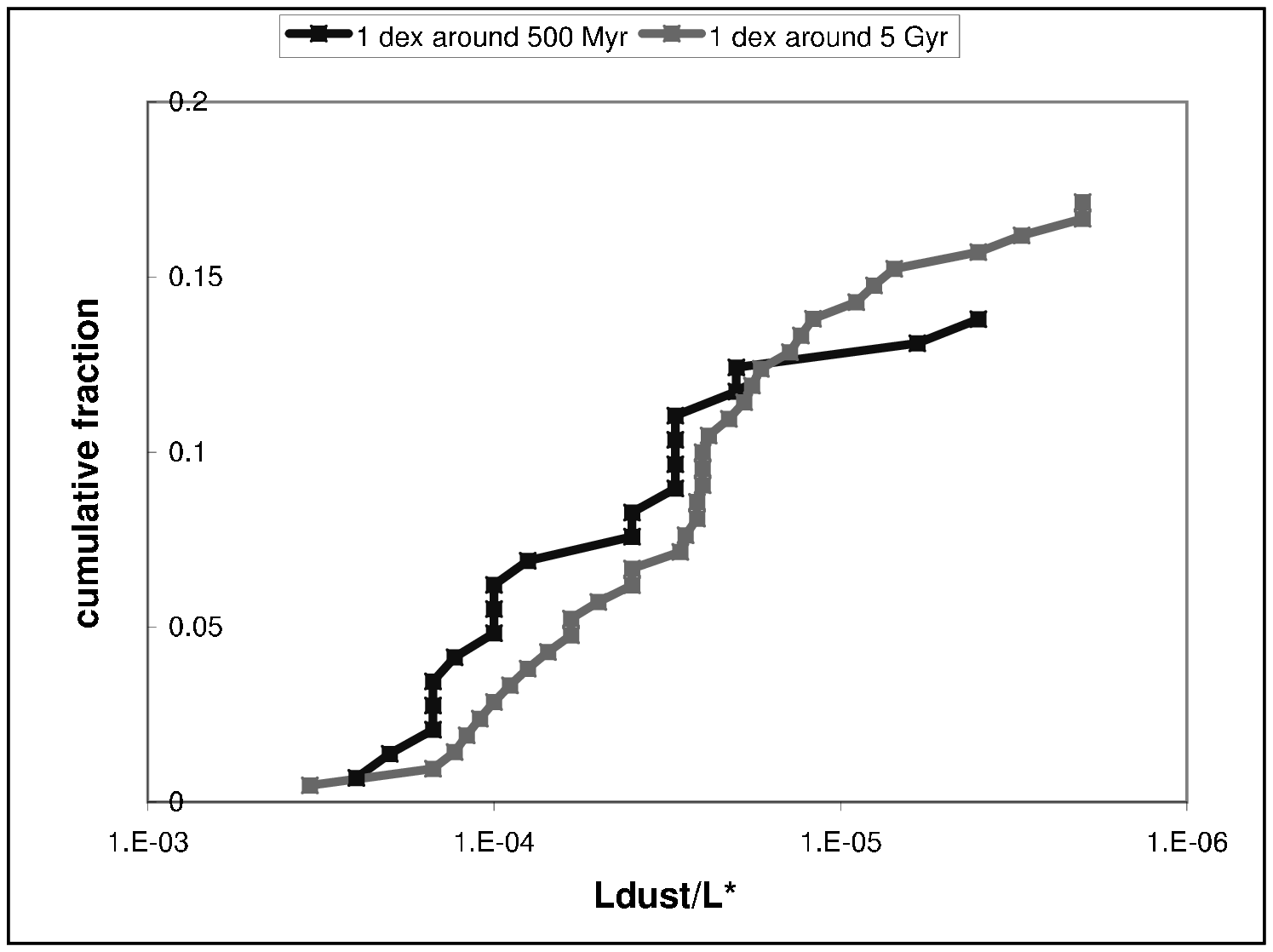}
 \caption{Cumulative $L_{dust}/L_*$ from far-infrared data for 
Sun-like stars, in 1 dex-wide bins at 500 Myr and 5 Gyr containing 
139 and 210 stars respectively. Data are from the age-selected 
FEPS project \citep{feps08} and volume limited surveys 
\citep{trilling,chas} with Spitzer plus an ISO search of the UMa 
moving group (see text). Some luminosities are lower limits 
(e.g. single-wavelength detection), and low values are 
incomplete, especially for ISO/FEPS at $\la 10^{-5}$, contributing 
$\approx 80, 20$~\% of targets in the young, old bins 
respectively. 
}
\end{figure}

One hypothesis is that high debris levels could be associated
with an early epoch while lower levels of dust do not have such
behaviour -- e.g. \citet{gw} noted that bright (IRAS-detected)
debris is marginally more common among G-dwarfs under 1~Gyr old
compared to older stars.  To re-assess this hypothesis, we
looked at far-infrared surveys of Sun-like stars and compiled
literature results into the same broad age bins around 500 Myr
and 5 Gyr (Figure~4).  Earlier observations with ISO of stars
in the 500~Myr-old UMa moving group \citep{spangler} were added
to improve the statistics in the young bin; there are three
far-infrared debris detections (for HD~125451, 139798, 184960)
among fourteen additional Sun-like stars observed. Figure~4
confirms that dust luminosity derived from far-infrared excess
is on average somewhat higher in the younger group. The
millimetre-systems also appear to be representative of this
luminous young group, with the four robust detections with good
dust temperature estimates yielding $L_{dust}/L_{star} \approx
5-8 \times 10^{-5}$ from the millimetre fluxes. 

\subsection{Comparison to models}

In general the Sun-like stars observed in millimetre surveys do
not show cold dust, nor are there debris detections later than
about 3~Gyr. This suggests that large late-evolving comet belts
are uncommon, such as those that could be stirred by formation of
distant dwarf planets at several-Gyr ages \citep{kb08}. While
there are examples of both very large discs
\citep{hd107146,liseau} and very old debris hosts
\citep{tauceti}, neither appears to be the norm among Sun-like
stars observed at millimetre wavelengths. 

To determine typical disc properties, we compare recent
(semi-)analytical models that predict the evolution of debris
with time under various scenarios
\citep{dominik,krivov,lohne,wyatt07,kb08}; see \citet{wyattrev}
for a summary. 
\begin{itemize}
\item{In {\it steady state} evolution, the fractional
luminosity of debris declines slowly as the parent population of
planetesimals is ground to dust that is blown out of the system.
The dust is brightest early on and then fades after a time that
is shortest in compact and/or massive discs, because of the
greater planetesimal collision rates. Examples of models that fit
the present data well \citep[Fig.~5]{wyattrev} are discs of 30~AU
radius with initial masses up to 1~M$_{\oplus}$ in colliding
bodies. If the majority are lower-mass systems (e.g. 
0.1~M$_{\oplus}$) then these are never detected as
$L_{dust}/L_{star}$ is always below $10^{-5}$, while the
highest-mass discs have $L_{dust}/L_{star}$ of a few $10^{-5}$ at
around 500~Myr but declining to $< 10^{-5}$ after around 3~Gyr.
These trends match most features of the data, including 
greater dust luminosity in the sub-Gyr stars.} 
\item{In contrast, in
models where the planetesimals are {\it stirred} by
gradually-forming larger bodies (from 1000~km sizes up to giant
planets), debris can brighten from an early low state and then
fade at a similar rate to steady-state models. These systems tend
to be more luminous than we observe unless the initial disc mass
is much less than in the Minimum Mass Solar Nebula
\citep[Fig.~5]{wyattrev}. For millimetre and far-infrared debris
detections to be made mainly for younger stars, discs should
generally be compact, e.g. 70~AU in radius \citep[Fig.~16]{kb08},
as larger discs will stay bright beyond 1~Gyr.} 
\item{Finally, in discs
stirred by {\it migration} of giant planets, there can be an
abrupt dust shower as these bodies cross an orbital resonance
with the comet belt, and very little debris later as the belt is
depleted. \citet{gomes} find that the migration of Jupiter and
Saturn could have stirred the Kuiper Belt thus causing the Late
Heavy Bombardment of the Earth at around 0.7~Gyr, while
\citet{thommes} simulate this for different planetary system
architectures and find equivalent events before $\sim 0.3$~Gyr. 
However, this is unlikely to be a general phenomenon as debris
would not be seen at all at few-Gyr ages when the comet belts
would be severely depleted. Further, the 70~$\umu$m excess is
predicted to increase only slightly during such events and the
dusty episode would last $\la 20$~Myr (Booth et al., 
submitted). Thus at most a few per cent of stars in the $\sim 
500$~Myr bin should be undergoing heavy bombardment episodes 
when observed, and even fewer if only up to $\approx 20$~\% of 
Sun-like stars host giant planets \citep{cumming}. }
\end{itemize}

\subsection{Disc populations}

In general, the low debris incidence at all ages and the mild
decline in luminosity at a few Gyr require initial planetesimal
discs of low mass, and of radius similar to the $\approx 50$~AU
of the Solar System. The stirred models have a distinctive
brightening at early times ($\sim 10^{7-8}$~yr) that the steady
state models do not, and the somewhat raised incidence of
70~$\umu$m excesses among FEPS stars at $10^{7.5-8}$ years
\citep[$\sim 20$~\%]{hillenbrand} could favour this class of
interpretations. Bright debris systems undergoing resonance
crossing in planetary migration are likely to be too few to be
observed as a distinct subset among current surveys. 

Resolved debris discs of Sun-like stars \citep{wyattrev,kalas}
are probably unusually large if most systems possess Solar
System-sized comet belts. For five precise (G2 V) Solar analogues
the fitted disc radii are $\sim 100-200$~AU \citep{krivov}, also
suggesting that the larger, more luminous, discs currently
dominate the detected population. To place the Sun's compact,
low-dust, Kuiper Belt more accurately within the context of
extrasolar debris systems requires deeper observations, and the
results here are preliminary given the small sample size. The
SCUBA survey was the first unbiased exploration of G-dwarfs in
the submillimetre, and the upcoming SCUBA-2 survey \citep{sunss}
will observe 100 G-dwarfs (as well as equal numbers of A, F, K
and M stars), and address the models and evolutionary timescales
in more detail. 

\section{Conclusions}

This small unbiased survey for debris around Solar analogues
suggests that cold bright dust-belts are rare, especially for
stars of typical few-Gyr ages. The majority of stars should begin
with insubstantial comet belts of dimensions similar to the Sun's
Kuiper Belt, with moderate dust luminosity only for about the
first Gyr of main-sequence lifetime. If these systems fade slowly
with time as predicted in models, counterparts of Sun-like age
should be detectable in the far-infrared to submillimetre with
near-future SCUBA-2 and Herschel observations. 

\section*{Acknowledgments}

The James Clerk Maxwell Telescope is operated by The Joint
Astronomy Centre on behalf of the Particle Physics and
Astronomy Research Council of the United Kingdom, the
Netherlands Organisation for Scientific Research, and the
National Research Council of Canada.  These data were
obtained under the long-term program M03BU30.


\bsp

\label{lastpage}

\end{document}